\documentclass[aps,prd,superscriptaddress,amsmath,groupedaddress,       
        reprint,showpacs,10pt]{revtex4-1}
\usepackage{revsymb}

\usepackage{amssymb,amsmath,amsthm} 
\usepackage{graphicx,ctable,booktabs}
\usepackage{url}
\usepackage[margin=1.25in]{geometry}
\begin{document}

\title{Comment on  Implications of new physics in the decays 
$B_c \to (J/\psi,\eta_c)\tau\nu$}

\author{C.~T.~H.~Davies}
\affiliation{SUPA, School of Physics and Astronomy, University of Glasgow, Glasgow, G12 8QQ, UK}

\author{C. McNeile}
\affiliation{Centre for Mathematical Sciences,
Plymouth University, Plymouth PL4 8AA, United Kingdom}

\date{\today}

\begin{abstract}
As part of a study BSM corrections to leptonic decays of the $B_c$
meson, Tran et al.~\cite{Tran:2018kuv} use the covariant
confining quark model (CCQM) to
estimate the matrix element of the 
pseudo-scalar curent between the vacuum and the $B_c$ meson. We note
that this matrix element can be determined using existing lattice QCD
results.
\end{abstract}

\maketitle

\section{Introduction}

The paper by Tran et al. ~\cite{Tran:2018kuv} discusses
Beyond the Standard Model (BSM) contributions 
to leptonic and semi-leptonic decays of the 
$B_c$ meson. This is a very topical calculation
because of the tantalizing hints that there are lepton symmetry
violations in various B and $B_c$ meson decays found by the 
LHCb collaboration~\cite{Aaij:2017tyk}.
To quantify the constraints from these analyses it is important 
to have reliable values for the operator matrix elements involved, 
with quantified uncertainties. 

\section{The pseudo-scalar matrix element}
\label{sec:psme}

\cite{Tran:2018kuv} considers a
Hamiltonian of corrections to the standard model:
\begin{equation}
{\cal H}_{eff} =
\frac{4 G_F V_{cb}}{\sqrt{2}}
(
{\cal O}_{V_L}
+ \sum_{X=S_i, V_i, T_L}  \delta_{l \tau}   X {\cal O_X}
)
\label{eq:Hnew}
\end{equation}
and works out the
phenomenology for the leptonic and semi-leptonic decays of the
$B_c$ meson.
The operators considered are:
\begin{eqnarray}
{\cal O}_{V_i} & = & (\overline{c} \gamma^\mu P_i b  )
 (\overline{l} \gamma_{\mu} P_L \nu_l ) ,  \\
{\cal O}_{S_i} & = & (\overline{c}  P_i b  )
 (\overline{l}  P_L \nu_l ), \\
{\cal O}_{T_L} & = & (\overline{c} \sigma^{\mu\nu} P_L b  )
 (\overline{l} \sigma_{\mu\nu} P_L \nu_l ) ,
\label{eq:BSMcontribution}
\end{eqnarray}
where
$\sigma_{\mu\nu} = i [\gamma_\mu , \gamma_\nu] / 2 $,
$P_L = (1 - \gamma_5 ) / 2 $, and 
$P_R = (1 + \gamma_5 ) / 2 $.

The delta function in the Hamiltonian in equation~\ref{eq:Hnew}
takes into account lepton flavor violation in this model.
The complex $X$ are the Wilson coefficients from the 
Beyond the Standard Model (BSM) theory. We note that 
there is no suppression of
the operators by the scale of the BSM physics, because the three
additional operators all have the same dimension as the 
operators in the standard model:

The leptonic decay constant of the $B_c$ meson, $f_{B_c}$:
\begin{equation}
\label{eq:fbc}
\langle 0 \mid \overline{c} \gamma_5 \gamma_{\mu} b \mid B_c \rangle =
f_{Bc}   p_{\mu} ,
\end{equation}
is used in the standard model calculation of the annihilation rate of the 
$B_c$ meson to leptons via a $W$ boson.
The additional operators in equation~\ref{eq:BSMcontribution} require
the  introduction of the pseudo-scalar matrix element
of the $B_c$ meson defined via
\begin{equation}
\langle 0 \mid \overline{c} \gamma_5 b \mid B_c \rangle = f_{Bc}^{P}(\mu)
  M_{bc}.
\end{equation}
The matrix element $f_{Bc}^{P}$ depends on the
renormalization scale $\mu$ in QCD. A physical result is obtained when it is
combined with the Wilson coefficient, which also depends on $\mu$, from the BSM theory.

The leptonic branching fraction of the $B_c$  meson is
\begin{multline}
{\cal B} (B_c \rightarrow \tau \nu) = 
\frac{G_F^2}{8 \pi} \mid V_{cb} \mid^2 \tau_{B_c}  m_{B_c} m_{\tau}^2  \\
\left( 1 - \frac{m_\tau^2}{m_{B_c}^2}  \right)^2 
f_{B_c}^2  
A_{BSM},
\end{multline}
where $A_{BSM}$  is

\begin{equation}
A_{BSM}  = \mid 1 - (V_R - V_L) + 
\frac{m_{B_c}}{m_\tau} \frac{f_{B_c}^{P} }{f_{B_c}} (S_R - S_L) 
\mid^2 .
\end{equation}

In the standard model $A_{BSM}$ = 1. If there are experimental
deviations of the leptonic decay of the $B_c$ meson from
the value in the standard model, then
the values of $f_{B_c}$ and $f_{B_c}^{P}$ are required to 
constrain values of the Wilson coefficients
$V_R$, $V_L$, $S_R$, and $S_L$, of the BSM theory.
The Wilson coefficients also contribute to semi-leptonic
decays of heavy light mesons, so additional constraints on them 
can be obtained.
This is an modern update of the experimental 
origins of the V-A theory in the standard model,
where experimental data was used to constrain the
interactions between quarks (see~\cite{Das:2009zzd} for example).
Although the leptonic decay of the $B_c$ meson has not been
observed experimentally, the constraints from a LEP1 
measurement allowed, Tran et al.~\cite{Tran:2018kuv}
to  put bounds on the $S_L$ and $S_R$ couplings.
~\cite{Tran:2018kuv,Ivanov:2016qtw}  
use a CCQM to estimate $f_{Bc}^{P}(\mu)$, although without giving 
a scale, $\mu$, at which it is determined.

\section{Lattice QCD results}
\label{sec:lattice}

The decay constant of the $B_c$ has been calculated in 
lattice QCD using two different approaches which give
results in good agreement~\cite{McNeile:2012qf, Colquhoun:2015oha}. 
The most accurate results comes from using the Highly 
Improved Staggered Quark (HISQ) formalism~\cite{Follana:2006rc}. 
In this formalism there is an exact partially conserved axial current
(PCAC)~\cite{Kilcup:1986dg} relation
\begin{equation}
\partial_{\mu} A_{\mu} =  (m_1  +  m_2)  P  \;\;.
\label{eq:PCACdefn}
\end{equation}
From the pseudoscalar matrix element times quark mass we 
can then obtain the matrix element of the temporal 
axial current (at zero spatial momentum) needed for 
eq.~(\ref{eq:fbc}) with absolute normalisation. 
This is done in~\cite{McNeile:2012qf} for heavy-charm 
pseudoscalar mesons for a range of heavy 
quark masses and values of the lattice spacing, $a$. This enables 
the heavy quark mass dependence of the heavy-charm 
decay constant to be mapped out in the continuum ($a \rightarrow 0$) 
limit and a result for $f_{B_c}$ to be obtained when 
the heavy quark mass corresponds to that of the $b$. 
The value obtained is 
\begin{equation}
\label{eq:fbcresult}
f_{B_c} = 0.427(6)(2) \,\mathrm{GeV}, 
\end{equation}
and a complete error budget is given in~\cite{McNeile:2012qf}. 

A completely different approach for $f_{B_c}$ based on the lattice 
discretisation of nonrelativistic QCD (NRQCD)~\cite{Lepage:1992tx} is given 
in~\cite{Colquhoun:2015oha}. There the matrix element 
of the temporal axial current is calculated directly but, since 
there is no PCAC relation on the lattice in this case, 
the current is matched to that of continuum QCD using lattice QCD perturbation 
theory through $\mathcal{O}(\alpha_s)$~\cite{Monahan:2012dq}. 
A result for $f_{B_c}$ of 0.434(15) GeV is obtained, where the uncertainty 
is dominated by that from lattice discretisation effects and 
systematic uncertainties in matching the current. Although 
the uncertainty is larger here than in the HISQ case, the 
agreement between the two results is confirmation of our 
understanding of the errors from the two approaches. 

Since, in the HISQ case~\cite{McNeile:2012qf}, the lattice PCAC 
relation was used to determine $f_{B_c}$, it is clear that 
we could also have determined $f_{Bc}^{P}(\mu)$.  Since $f_{Bc}^P(\mu)$ 
runs with $\mu$ it is much more convenient to determine it in combination 
with quark masses. The PCAC relation, eq.~(\ref{eq:PCACdefn}) on the lattice 
yields the following relationship between $f_{B_c}$ and $f_{Bc}^P$:
\begin{equation}
(m_b+m_c)f_{Bc}^{P} = M_{B_c} f_{Bc} .
\label{eq:pcaclatt}
\end{equation}
Here $m_b$ and $m_c$ are the bare lattice quark masses. 
Since both sides of this equation are scheme- and scale-invariant, 
we can instead apply this relationship in the continuum 
using the continuum results for $f_{B_c}$ obtained from 
lattice QCD calculations. 

Then
\begin{equation}
f_{Bc}^{P}(\mu) = \frac{M_{B_c} f_{Bc} }{m_b(\mu) + m_c(\mu)}, 
\label{eq:pcacdecay}
\end{equation}
where $m_b(\mu)$ and $m_c(\mu)$ are the bottom and charm 
quark masses
at the scale $\mu$ in a standard continuum 
scheme, such as $\overline{\mathrm{MS}}$.
The quark masses are also most conveniently and accurately obtained 
from lattice QCD calculations, see for example~\cite{McNeile:2010ji}. 

We use results from~\cite{McNeile:2010ji} for the quark masses 
in the $\overline{\mathrm{MS}}$ scheme at a standard scale of 
3 GeV,
$\overline{m}_c$(3 GeV, $n_f$=4) = 0.986(6) GeV, $m_b/m_c$ = 4.51(4), 
$f_{B_c}$ from~\cite{McNeile:2012qf} (0.427(6) GeV)
and $M_{B_c}$ = 6.274(1) GeV 
from experiment~\cite{Aaij:2016qlz}. 
This gives, in the $\overline{\mathrm{MS}}$ scheme 
\begin{equation}
\label{eq:fbcpresult}
\overline{f}_{Bc}^P (\mbox{3 GeV}) = 0.493(9)  \,\,\mathrm{GeV}
\end{equation}
where the uncertainty is dominated by that from the lattice 
QCD result for $f_{B_c}$. 
The result for $\overline{f}_{Bc}^{P}$ 
can be run to different values of $\mu$ using the 
inverse of the running of the $\overline{\mathrm{MS}}$ quark 
mass~\cite{Chetyrkin:1997dh, Vermaseren:1997fq}. 

The result for $f_{B_c}$ computed using the CCQM~\cite{Tran:2018kuv} 
of 0.489 GeV is 15\% larger than that obtained from the 
lattice QCD results discussed above. The systematic uncertainty from 
using the CCQM is estimated in~\cite{Tran:2018kuv}
as 10\%. 
The result given in~\cite{Tran:2018kuv} for $f_{B_c}^{P}$ of 0.645 GeV 
is hard to interpret or compare to the lattice QCD values since 
no scheme or scale for it is given. 
Lattice QCD results for the form factors of $B_c$ semileptonic 
decay to charmonium states are as yet preliminary~\cite{Colquhoun:2016osw} 
but will provide a further point of comparison in future.  

\section{Conclusions}

Weak decays of the $B_c$ meson provide exciting opportunities for 
constraining new physics as growing datasets from LHC, along with new 
analyses, become 
available~\cite{Tran:2018kuv}. 
The theoretical input to this of hadronic parameters such 
as decay constants and form factors for the $B_c$ 
need to be firmly based on `first-principles' approaches to 
QCD, such as lattice QCD. This allows not only the result to be 
given but also a well-motivated uncertainty on its value. 
To this end, we collect here existing lattice 
QCD results, with their associated 
uncertainty, for the $B_c$ 
decay constant and we derive from them a value for the pseudoscalar current 
matrix element. 

\begin{acknowledgments}
Our calculations were done on the Darwin Supercomputer as part of
STFC's DiRAC facility jointly funded by STFC, BIS and the Universities
of Cambridge and Glasgow.  This work was funded by STFC.
\end{acknowledgments}


\end{document}